\documentstyle[12pt]{article}
\begin{document}
\begin{center}
\section*{Transverse momentum dependence in the perturbative \\
calculation of pion form factor \\}
\end{center}
\begin{center}
Fu-guang Cao \footnote{E-mail address: caofg@bepc3.ihep.ac.cn}
and Tao Huang\\
CCAST (World Laboratory), P.O.Box 8730, Beijing, 100080 China\\
Institute of High Energy Physics, Academia Sinica,
P.O.Box 918, Beijing, 100039, China\\
\end{center}
\vskip 1.cm
\begin{center}
{\bf Abstract}
\end{center}
\vskip 0.5cm

\begin{minipage}{14.cm}
By reanalysing transverse momentum dependence
in the perturbative calculation of pion form factor
an improved expression of pion form factor which takes
into account the
transverse momentum dependence
in hard scattering amplitude
and intrinsic transverse momentum dependence associated with
pion wave functions is given to
leading order, which is available for momentum transfers of the
order of a few GeV as well as for $Q \rightarrow \infty$.
Our scheme can be extended to evaluate the contributions 
to the pion form factor beyond leading order.
\end{minipage}

\vskip 1.0cm
\begin{flushleft}
PACS number(s): 12.38.Bx, 13.40.Gp, 14.40.Aq
\end{flushleft}


\newpage
\section{Introduction}
Exclusive processes in perturbative
quantum chromodynamics (PQCD) were first studied by Brodsky
and Lepage \cite{Brodsky} many years ago.
Recently, the studies of the exclusive processes at experimentally
accessible momentum transfers in the framework of
PQCD have received much interest.
The statement \cite{Isgur,Smith}
that the applicability of PQCD to the exclusive 
processes at experimentally accessible momentum transfers
is questionable has been challenged [4-11].
Huang and Shen \cite{Shen} pointed out that the applicability of 
PQCD to form factor is questionable only as momentum
transfer $Q^2 \leq 4$ GeV$^2$ through the study on the pion form 
factor by reanalyzing the contributions from 
end-point regions.
Botts and Sterman \cite{Botts} proposed a formulation in which 
both Sudakov and nonleading logarithmic corrections to independent
(Landshoff) scatterings \cite{Land} of valence quarks
are organized systematically.
Li and Sterman \cite{Li} gave out a modified expression for the
pion form factor by taking into account the customarily
neglected partonic transverse momentum as well as 
Sudakov correction.
They reached a similar conclusion as \cite{Shen}:
PQCD begins to be self-consistent at about $Q \sim 20 \Lambda_{QCD}$.
Jakob and Kroll \cite{Jakob} demonstrated that for momentum
transfers of the order of a few GeV the intrinsic
transverse momentum dependence of wave function 
leads to a substantial suppression of the perturbative
contributions, which should be considered besides Sudakov
suppression.
The base of the most previous discussions [5-11] is
the formalism in Ref. \cite{Botts} which is suitable for studying
the large $Q$ region since it sets $b \rightarrow 0$
in the integral of the wave function in respecting the
intrinsic transverse momentum dependence of wave function.
In this paper we re-analyse the  PQCD calculation for the 
pion form factor at momentum transfers of the order of
a few GeV.
We will give out an improved 
expression for the pion form factor which takes
into account the
transverse momentum dependence in the gluon propagator
as well as in the fermion propagator
in the hard scattering amplitude $T_H$
and the intrinsic transverse momentum dependence associated with
the pion wave function.
This expression is available for momentum transfers 
of the order of a few GeV as well as for $Q \rightarrow \infty$
since it dose not make the approximation $b \rightarrow 0$.
The formalism of Ref. \cite{Jakob}
is just an approximate expression
in respecting the intrinsic
transverse momentum dependence. This approximation brings sizeable
effect on the numerical prediction for the 
pion form factor in the momentum transfer $Q \sim$ a few GeV region.
It is also found that the 
transverse momentum dependence of the fermion propagator in 
$T_H$ leads to a mild reduction of the prediction for 
the pion form factor 
in the same momentum transfer region.
The remainder of the paper is organized as follows.
Sect. 2 reviews LI-Sterman's formalism for the pion form factor.
In Sect. 3, we discuss the transverse momentum dependence in
pion wave function beside the one in the hard scattering amplitude $T_H$
in perturbative calculation.
In Sect. 4, we do our numerical calculations.
Finally, in Sect. 5, we give a summary.


\section{Brief review of Li-Sterman's formalism}
Taking into account 
the transverse momenta ${\bf k}_{T}$ that 
flow from the wave functions through the hard scattering
leads to a factorization form with two wave
functions $\psi(x_{i},{\bf k}_{T_{i}})$ corresponding
to the external pions, combined with a new hard-scattering
function $T_{H}(x_{1},x_{2},Q,{\bf k}_{T_{1}},{\bf k}_{T_{2}})$,
which depends in general on transverse as well as longitudinal
momenta \cite{Li},
\begin{eqnarray}
F_{\pi}(Q^{2})&=&\int_0^1 dx_{1}dx_{2}\int \frac{d^{2}{\bf k}_{T_{1}}}
{16 \pi^3} \frac {d^{2}{\bf k}_{T_{2}}}{16 \pi ^3}
\psi(x_{1},{\bf k}_{T_{1}}) \nonumber\\
&\times& T_{H}(x_{1},x_{2},Q,{\bf k}_{T_1},{\bf k}_{T_2})
\psi(x_{2},{\bf k}_{T_{2}}).
\label{fk}
\end{eqnarray}
In this form, both soft and collinear logarithmic enhancements are
factorized into the functions $\psi$.
At the lowest order, $T_H$ reads \footnote{ A frame with 
$q_T=0$
has been adopted to obtain this expression of $T_H$ in \cite{Li}.
The expressions of $T_H$ with different momentum assignment
will be analysed in detail \cite{CFG}.}
\begin{eqnarray}
T_{H}(x_{1},x_{2},Q,{\bf k}_{T_1},{\bf k}_{T_2})=
\frac{16\pi C_{F} \alpha_{s}(\mu) x_1 Q^2}
{ (x_1 Q^2 + {\bf k}_{T_1}^2)\left( x_{1}x_{2}Q^{2}
+({\bf k}_{T_{1}}+{\bf k}_{T_{2}})^{2} \right)}.
\label{THgf}
\end{eqnarray}
Neglecting the transverse momentum dependence in the fermion 
propagator, Eq. (\ref{THgf}) becomes 
\begin{eqnarray}
T_{H}(x_{1},x_{2},Q,{\bf k}_{T_{1}},{\bf k}_{T_{2}})=
\frac{16\pi C_{F} \alpha_{s}(\mu)}
{x_{1}x_{2}Q^{2}
+({\bf k}_{T_{1}}+{\bf k}_{T_{2}})^{2}}.
\label{THg}
\end{eqnarray}

The next step is to re-express Eq. (\ref{fk}) in terms of the 
Fourier transformation variables in the transverse configuration space
\cite{Li}. Observing that $T_H$ in Eq. (\ref{THg})
depends on only a single 
combination of the transverse momenta 
$({\bf k}_{T_1}+{\bf k}_{T_2})$,
the Fourier transformation
of Eq. (\ref {fk}) involves only a single integral 
of Fourier transform variable $b$,
\begin{eqnarray}
F_{\pi}(Q^{2})=\int_0^1 dx_{1}dx_{2}\frac{d{\bf b}}{(2\pi)^{2}}
 {\varphi}(x_{1},{\bf b},\mu)
 {T}_{H}(x_{1},x_{2},Q,{\bf b},\mu)
 {\varphi}(x_{2},{\bf b},\mu).
\label{fbcao}
\end{eqnarray}
The wave functions in $b$-space,
${\varphi}(x,{\bf b},\mu)$
take into account
an infinite 
summation of higher-order effects associated with the elastic
scattering of valence partons, which gives out Sudakov 
suppression to the large-$b$ and small-$x$ regions. 

The asymptotic behavior of $ {\varphi}(x,{\bf b},\mu)$
at large $Q^2$ has been obtained in Ref. \cite{Botts}
\begin{eqnarray}
 {\varphi}(x,b,\mu)=\exp\left[ -s(x,b,Q)-s(1-x,b,Q)
-2 \int_{1/b}^{\mu}\frac{d\bar{\mu}}{\bar{\mu}}\gamma_{q}
(g(\bar{\mu}))\right]\times  {\phi}\left(x,\frac{1}{b}\right),
\label{varphi}
\end{eqnarray}
where $\gamma_{q}=-\alpha_{s}/\pi$ is the quark anomalous dimension in
the axial gauge.
$s(\xi,b,Q)$ is Sudakov exponential factor,
which reads \cite{Botts,cao}
\begin{eqnarray}
s(\xi,b,Q)&=&\frac{A^{(1)}}{2\beta_{1}}\hat{q}~
\ln\left(\frac{\hat{q}}{-\hat{b}}\right)+\frac{A^{(2)}}{4\beta_{1}^{2}}
\left(\frac{\hat{q}}{-\hat{b}}-1\right)-\frac{A^{(1)}}{2\beta_{1}}
(\hat{q}+\hat{b}) \nonumber\\
&-&\frac{A^{(1)}\beta_{2}}{4\beta_{1}^{3}}\hat{q}
\left[\frac{\ln(-2\hat{b})+1}{-\hat{b}}-\frac{\ln(-2\hat{q})+1}
{-\hat{q}}\right] \nonumber\\
&-&\left(\frac{A^{(2)}}{4\beta_{1}^{2}}-
\frac{A^{(1)}}{4\beta_{1}}\ln(\frac{1}{2}{\rm e}^{2\gamma-1})\right)
\ln\left(\frac{\hat{q}}{-\hat{b}}\right) \nonumber\\
&+&\frac{A^{(1)}\beta_{2}}{8\beta_{1}^{3}}
\left[\ln^{2}(2\hat{q})-\ln^{2}(-2\hat{b})\right],
\label{smalls}
\end{eqnarray}
where 
\begin{eqnarray}
\hat{q}&=&\ln[\xi Q/(\sqrt{2}\Lambda)],
 ~~~~ \hat{b}=\ln(b\Lambda),\nonumber\\
\beta_{1}&=&\frac{33-2n_{f}}{12}, ~~~~~  \beta_{2}=\frac{153-19
n_{f}}{24}, \nonumber\\
A^{(1)}&=&\frac{4}{3},~~~~~~ A^{(2)}=\frac{67}{9}-\frac{1}{3}\pi^{2}
-\frac{10}{27}n_{f}+\frac{8}{3}\beta_{1}\ln(\frac{1}{2}{\rm e}^{\gamma}).
\end{eqnarray}
$n_{f}$ is the number of quark flavors. $\gamma$
is the Euler constant.
The factor ${\rm e}^{-s(\xi,b,Q)}$ which induces an enhancement in
small-$b$ regions has been set to unity 
whenever $\xi \le {\sqrt 2}/(bQ)$.
The coefficients in Eq. (\ref {smalls}) are different from Ref. \cite{Li}
since there are some algebraically
mistakes in the previous expression of $s(\xi,b,Q)$, which
has been pointed in Ref. \cite{cao}.
${\phi}\left(x,\frac{1}{b}\right)$ in Eq. (\ref {varphi})
is a ``soft" wave function 
calculated with gluons of transverse momentum $k_T$ $\leq$ $1/b$,
\begin{eqnarray}
\phi(x,1/b)=\int_{k_T \leq 1/b} \frac{d^{2}{\bf k}_T}{16 \pi^3} 
\psi(x,{\bf k}_T).
\label{wbotts}
\end{eqnarray}

Neglecting the $b$-dependence of the function $\phi(x,1/b)$ and
performing Fourier transformation for Eq. (\ref{THg}),
the expression Eq. (\ref{fk}) becomes \cite{Li}
\begin{eqnarray}
F_{\pi}(Q^{2})&=&16\pi C_{F}\int_{0}^{1}d x_{1}d x_{2}
\int_{0}^{\infty}b~db \alpha_{s}(t)K_{0}(\sqrt{x_{1}
x_{2}}Qb)\nonumber\\
&\times& \exp \left(-S(x_{1},x_{2},Q,b,t)\right)
\phi(x_1) \phi(x_2),
\label{ffli}
\end{eqnarray}
where $\phi(x)$ is defined as
\begin{eqnarray}
\phi(x)=\int \frac{d^2 {\bf k}_T}{16 \pi ^3} \psi(x,\bf{k}_T),
\end{eqnarray}
$K_0$ is the modified Bessel function of order zero,
and
\begin{eqnarray}
S(x_{1},x_{2},Q,b,t)=
\left[\sum_{i=1}^{2}\left(s(x_{i},b,Q)+
s(1-x_{i},b,Q)\right)-\frac{2}{\beta_{1}}\ln\frac{\hat{t}}
{-\hat{b}}\right].
\end{eqnarray}

\section{$k_T$-dependence in perturbative calculation}
Following Ref. \cite{Botts}, we find that $\phi(x,1/b)$ 
in Eq. (\ref{wbotts})
should be replaced by the amplitude
\begin{eqnarray}
{\varphi}^{(1/b)}(x,b)=\int_{k_T \leq 1/b} \frac{d^{2}{\bf k}_T}
{16 \pi^3} 
~{\rm e}^{i {\bf k}_T\cdot {\bf b}}
\psi(x,{\bf k}_T).
\label{wcao}
\end{eqnarray}
Through the requirement that ${\rm e}^{i {\bf k}_T \cdot {\bf b}} \approx 1$,
as $Q \rightarrow \infty$ ($b \rightarrow 0$).
Ref. \cite{Botts} proposed $ {\varphi} ^{(1/b)}(x,b)$ can be
expressed approximately by $ {\phi}(x,1/b)$.
For simplicity, Ref. \cite{Li} neglected the $b$-dependence of the
function $ {\varphi}^{(1/b)}(x,b)$, namely neglected 
the intrinsic transverse
momentum dependence of wave function. 
Then $ {\varphi}^{(1/b)}(x,b)$
can be pulled out of the $b$-integral.
Jakob and Kroll \cite{Jakob} improved this approximation and 
proposed to take into account 
the intrinsic transverse momentum
dependence by a function $ {\psi}(x,b)$ which is the
Fourier transformation of wave function,
\begin{eqnarray}
 {\psi}(x,b)=\int_{k_T\leq \infty} \frac{d^{2}{\bf k}_T}{16 \pi^3}
~{\rm e}^{i {\bf k}_T\cdot {\bf b}}
\psi(x,{\bf k}_T).
\label{wjakob}
\end{eqnarray}
It can be seen
that $ {\psi}(x,b)$ is just an approximate
expression for $ {\varphi}^{(1/b)}(x,b)$, i.e. 
$\psi (x,b)=\varphi^{(\infty)}(x,b)$.
When $b \rightarrow 0$ (namely $Q \rightarrow \infty$) they
are consistent with each other. When $Q$ is the order of a few GeV,
the difference may be sizable.
The difference between $ {\varphi}^{(1/b)}(x,b)$ 
and $ {\psi}(x,b)$ can be
expressed by a function $D(x,b)$,
\begin{eqnarray}
D(x,b)=\int_{1/b}^{\infty} \frac{d^{2}{\bf k}_T}{16 \pi^3}
~{\rm e}^{i {\bf k}_T\cdot {\bf b}}
\psi(x,{\bf k}_T),
\label{D}
\end{eqnarray}
which increases as $b$ becomes large.
Eq. (\ref {wjakob}) enlarges the upper limit of the integral
Eq. (\ref {wcao})
from $1/b$ to $\infty$, which corresponds to evaluate the 
contributions from the perturbative tail of wave
function once again.
Sudakov form factor provides much more suppression for
large $Q$ than for small $Q$. Thus substituting $\psi(x,b)$
for $\varphi^{(1/b)}(x,b)$
does not effect the pion form factor
for large Q.
As $Q$ is the order of a few GeV
one should investigate the effects of this substitution.

The Fourier-transformed hard-scattering amplitude 
from Eq. (\ref{THg}) reads
\begin{eqnarray}
{T}_H(x_1,x_2,Q,b,\mu)=16 \pi \alpha_s(\mu) C_F 
K_0\left(\sqrt {x_1 x_2} Q b \right).
\label{thb}
\end{eqnarray}
The renormalization group applied to $T_H$ gives 
\begin{eqnarray}
 {T}_H(x_1,x_2,Q,b,\mu)=\exp \left[-4\int_\mu^t \frac {d \bar{\mu}}
{\bar {\mu}} \gamma_q(g(\bar{\mu}))\right]
\times  {T}_H(x_1,x_2,Q,b,t).
\label{thbrgroup}
\end{eqnarray}
The variable $t$ was taken as 
$t=\max\left(\sqrt{x_{1} x_{2}} Q,1/b \right)$ and a cut-off on 
the running coupling constant 
($\alpha_s \le 0.7$) for large-$b$ region was made to
guarantee PQCD to be self-consistent in Ref. \cite{Li}.
However, in the regions of small $x_1 x_2 Q^2$ and large $b$,
the nonperturbative contributions maybe important. For example,
the multi-gluon exchange can occur between quark and antiquark
and the transverse momentum 
intrinsic to the bound
state wave-functions flows through all the propagators.
Ref. \cite{BHL} suggested a frozen $\alpha_s$
to take into account these effects.
Instead of the cut-off of $\alpha_s$ in the $b$-space, Ref. \cite{cao}
suggested that the coupling constant is
frozen at $b \sim 1/(\sqrt{\langle{\bf k}_{T}^{2}\rangle})$
and the variable $t$ 
is taken as
\begin{eqnarray}
t=\max\left(\sqrt{x_{1} x_{2}} Q,1/b_F \right),
\end{eqnarray}
where
\begin{eqnarray}
b_F=\left \{
	\begin{array}{cl}
	b & \mbox{if $1/b \geq \sqrt{\langle{\bf k}_{T}^{2}\rangle}$}\\
	1/\sqrt{\langle{\bf k}_{T}^{2}\rangle}~~~~~~~~~ & 
	\mbox{if $1/b< \sqrt{\langle{\bf k}_{T}^{2}\rangle},$} 
	\end{array}
	\right. 
\end{eqnarray}
and $\sqrt{\langle{\bf k}_{T}^{2}\rangle}$ is 
the average transverse momentum of the pion.
In this way, the perturbative contributions
to the pion form factor  can be calculated from the present energy
with a reasonable $\alpha_{s}$.
It should be emphasized that the average transverse momentum
$\sqrt{\langle{\bf k}_{T}^{2}\rangle}$  
is determined definitely by the hadronic wave function.

Combining Eqs. (\ref {fbcao}), (\ref{varphi}), (\ref{wcao}),
(\ref{thb}) 
and (\ref{thbrgroup}), we have
\begin{eqnarray}
F_{\pi}(Q^{2})&=&16\pi C_{F}\int_{0}^{1}d x_{1}d x_{2}
\int_{0}^{\infty}b~db \alpha_{s}(t)K_{0}(\sqrt{x_{1}
x_{2}}Qb)\nonumber\\
&\times& \exp \left(-S(x_{1},x_{2},Q,b,t)\right)
 {\varphi}^{(1/b)}(x_{1},b)
 {\varphi}^{(1/b)}(x_{2},b),
\label{ff}
\end{eqnarray}
where
\begin{eqnarray}
S(x_{1},x_{2},Q,b,t)=
\left[\sum_{i=1}^{2}\left(s(x_{i},b,Q)+
s(1-x_{i},b,Q)\right)-\frac{2}{\beta_{1}}\ln\frac{\hat{t}}
{-\hat{b}}\right].
\end{eqnarray}
Eq. (\ref{ff}) is an improved expression for the pion form factor.
From our formalism, it can be found that:
(a) neglecting the transverse momentum dependence associated with 
the wave function,
Eq. (\ref{ff}) becomes the expression in Ref. \cite{Li} 
(See Eq. (\ref{ffli}));
(b) approximating $\varphi^{(1/b)}(x,b)$
with  $\psi(x,b)$, Eq. (\ref{ff}) becomes the expression in 
Ref. \cite{Jakob},
\begin{eqnarray}
F_{\pi}(Q^{2})&=&16\pi C_{F}\int_{0}^{1}d x_{1}d x_{2}
\int_{0}^{\infty}b~db \alpha_{s}(t)K_{0}(\sqrt{x_{1}
x_{2}}Qb)\nonumber\\
&\times& \exp \left(-S(x_{1},x_{2},Q,b,t)\right)
\psi(x_{1},b)
\psi(x_{2},b),
\label{ffjakob}
\end{eqnarray}
where $\psi(x,b)$ is the Fourier transformation of the
wave function (Eq. (\ref{wjakob})).

The $k_T$-dependence in the fermion
propagator contributes to $T_H$ a factor $\frac{x_1 Q^2}
{x_1 Q^2 + {\bf k}_{T_1}^2}$ which involves only a 
single transverse momentum corresponding to one 
in the external pion (see Eqs. (\ref{THgf}) and (\ref{THg}))
and this factor makes
the Fourier transformation for $T_H$
involve multi-$b$-integrals\cite{Li2}.
Although the $k_T$-dependence in this factor is linear
rather than quadratic in the $x$'s, it may bring some effects
at the end-point region $x_1 \sim 0$.
We can consider this factor and keep the simplification
of the Fourier transformation for $T_H$ at the same time 
by combining it with the corresponding wave function
to define a new ``wave function",
\begin{eqnarray}
\tilde{\psi}(x_1,{\bf k}_{T_1})=\frac{x_1 Q^2}
{x_1 Q^2 + {\bf k}_{T_1}^2} \psi(x_1,{\bf k}_{T_1}).
\label{tldw1}
\end{eqnarray}
Substituting $\tilde{\psi}$ for $\psi$ in Eq. (\ref {fk}),
the transverse momentum
dependence of the fermion propagator in $T_H$ 
can be taken into account easily
in the perturbative calculations for the pion form factor.

\section{Numerical calculations}
We adopt two models of the pion wave function: 
(a) the BHL wave function \cite{BHL}
\begin{eqnarray}
\psi^{(a)}(x,{\bf k}_T)=A~\exp\left[-\frac{{\bf k}_{T}^{2}+m^{2}}
{8 \beta^{2} x(1-x)}\right],
\end{eqnarray}
where \cite{MABQ} $A = 32$ GeV$^{-1}$, $\beta = 0.385$ GeV,
$m = 289$ MeV and $\sqrt{\langle{\bf k}_{T}^{2}\rangle} = 356$ MeV;
(b) the CZ-like wave function [16-18]
\begin{eqnarray}
\psi^{(b)}(x,{\bf k}_T)=A~(1-2x)^2~\exp\left[-\frac{{\bf k}_{T}^{2}+m^{2}}
{8 \beta^{2} x(1-x)}\right],
\end{eqnarray}
where \cite{MABQ} $A = 136$ GeV$^{-1}$, $\beta = 0.455$ GeV,
$m = 342$ MeV and $\sqrt{\langle{\bf k}_{T}^{2}\rangle} = 343$ MeV.

In order to discuss the effect of transverse momentum dependence
associated with wave function, first, we neglect
the $k_T$-dependence
of the fermion propagator in $T_H$.
Fig. 1 compares the behaviors of functions $\varphi^{(1/b)}(x,b)$,
$\psi(x,b)$ and $\phi(x,1/b)$. 
All of them suppress the the contributions
from large-$b$ region, but the suppression behaviors
are different in quantity.
In the $b\sim 0$ region, $\phi(x,1/b)$ and $\psi(x,b)$ are good
approximation for $\varphi^{(1/b)}(x,b)$, while in 
the large-$b$ region that is questionable.
Sudakov form factor suppresses the contributions from large-$b$ 
region in a mild way
for small momentum transfer $Q$ than for large $Q$.
Thus it is not a good approximation
with $\psi(x,1/b)$ to respect the intrinsic $k_T$-dependence
of the wave functions
for small $Q$, since much more contributions come
from large-$b$ region.
Numerical evaluations of $F_\pi$ through Eqs. (\ref{ffli}), (\ref{ff})
and (\ref{ffjakob}), using the BHL and CZ-like wave functions,
which are plotted in Figs. 2 and 3 confirm the
observation made in Fig. 1. 
Comparing with the original PQCD prediction \cite{Brodsky}
of the pion form
factor (namely neglecting the $k_T$-dependence in $T_H$ as well as 
in wave functions;
the dotted line), Sudakov form factor gives
a suppression effect to $F_\pi$, which was first
pointed out by Brodsky and Lepage \cite{Brodsky}.
Comparing with the result obtained from Eq.(\ref{ffli}) (namely
neglecting the intrinsic $k_T$-dependence of wave function;
the dash-dotted line),
$\varphi^{1/b}(x,b)$ (the solid line) 
suppresses the contribution from PQCD
by about $50\%$, while $\psi(x,b)$ (the dashed line)
suppresses by about $30\%$ at
$Q=2$GeV for the BHL wave function. 
The corresponding quantities 
are $55\%$ and $35\%$ respectively
in the case of the CZ-like wave function.
The intrinsic $k_T$-dependence of the wave function
provides additional substantial suppression for $F_\pi$
besides Sudakov form factor.
$\varphi^{1/b}(x,b)$ suppresses the contribution from PQCD
more strongly than $\psi(x,b)$ does.
The numerical predictions of $F_\pi$ 
obtained from $\varphi^{1/b}(x,b)$ and
$\psi(x,b)$ are different 
in the momentum transfer $Q \sim$ a few GeV region.

We evaluate the effect of the $k_T$-dependence
in the fermion propagator also.
It leads to a small reduction 
of the prediction for $F_\pi$ by about $10\%$ for $Q \sim 3$ GeV,
which is coincides with Ref. \cite{Li2}.

\section{Summary}
In this paper we re-analyze the 
transverse momentum dependence in the
perturbative calculation of pion form factor
at momentum transfers of the order of a few GeV.
We give out an improved expression for the pion form factor which takes
into account gluon propagator as well as
fermion propagator 
transverse momentum dependence in the hard scattering amplitude
and intrinsic transverse momentum dependence associated with
pion wave functions.
It is found that the previous approach 
is just an approximate expression
in respecting the
transverse momentum dependence associated with wave functions.
This approximation brings sizeable
effect on the numerical predictions for the 
pion form factor in the momentum transfer $Q \sim$ a few GeV region.
It is also found that the 
transverse momentum dependence of fermion propagator in 
$T_H$ leads to a mild reduction of the prediction for 
the pion form factor 
in the same momentum transfer region.

We would like point out one more times that our
formalism is available for momentum transfer of the 
order of a few GeV as well as for $Q \rightarrow \infty$,
and our scheme can be extended to evaluate the higher order and
higher helicity contributions for the pion form factor.
The more studies on the $k_T$-dependence in the
hard scattering amplitude $T_H$ are in proceeding.

\begin{center}
{\bf Acknowledgements}
\end{center}
We are grateful to S. Brodsky and H. N. Li
for useful discussions.

\newpage
\parindent=0pt
\section*{Figure Captions}
\begin {description}
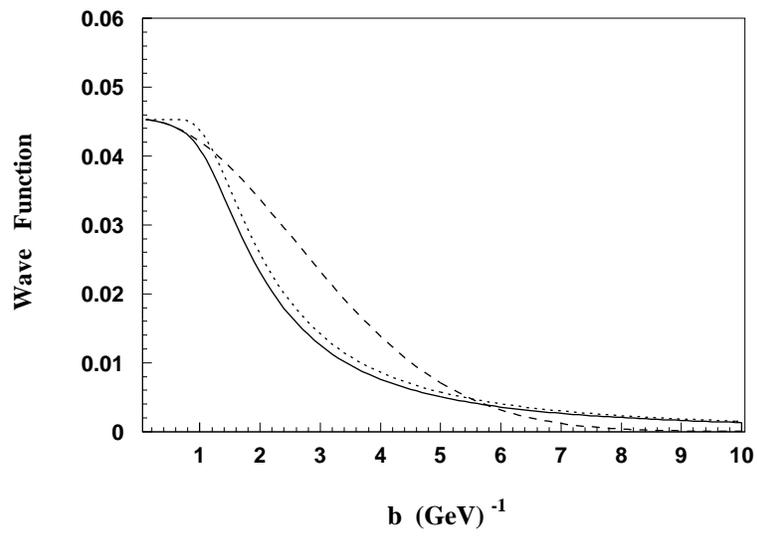
\item [Fig. 1.]
The functions $\varphi^{(1/b)}(x,b)$ (solid line),
$\psi(x,1/b)$ (dashed line) and 
$\phi(x,b)$ (dotted line)
which are adopted to respect the intrinsic transverse
momentum dependence.\\
$x=0.5$, $Q^2=4$ GeV $^2$, for the BHL wave function.

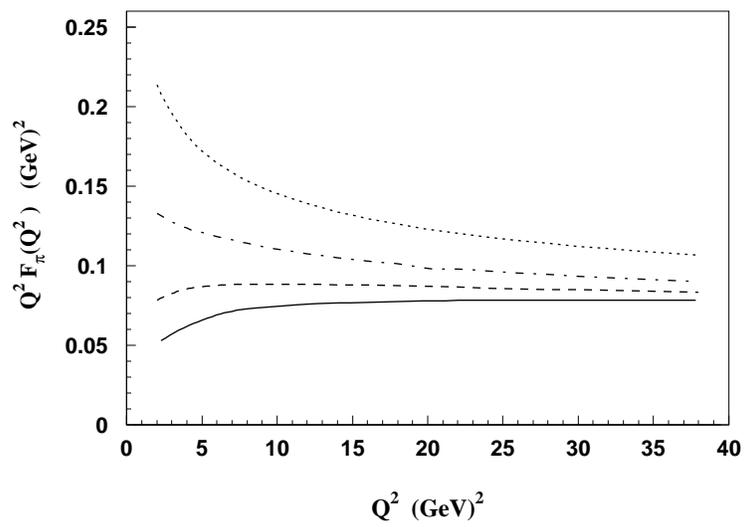
\item [Fig. 2.] 
The pion form factor calculated with the BHL wave function.
The solid and dashed lines are for
$\varphi^{(1/b)}(x,b)$ and
$\psi(x,1/b)$ 
respectively. The dash-dotted line is obtained
by neglecting the intrinsic transverse momentum dependence
in wave functions.
The dotted line is evaluated by neglecting $k_T$-dependence
in both $T_H$ and wave functions.
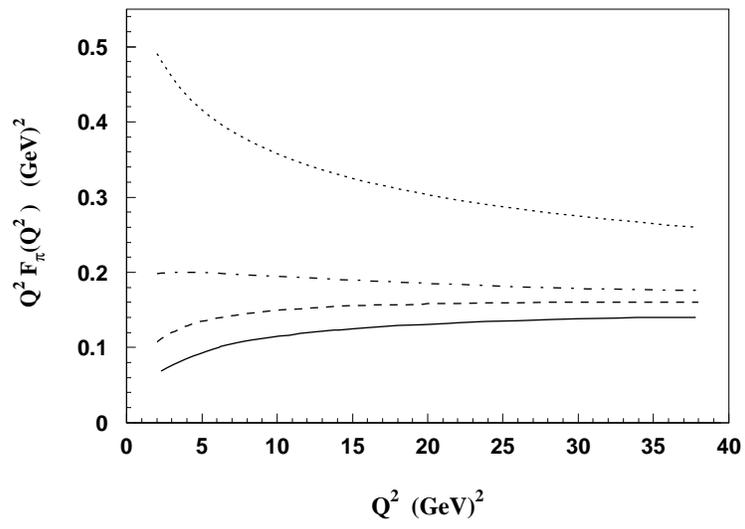
\item [Fig. 3.] 
Similar to Fig. 2.
The pion form factor calculated with the CZ-like wave function.

\end {description}


\newpage

\begin{thebibliography}{99}
\bibitem{Brodsky}
G.P. Lepage and S.J. Brodsky, Phys. Lett. {\bf 43} 545 (1979);
Phys. Rev. D {\bf 22} 2157 (1980).
\bibitem{Isgur}
N. Isgur and C.H. Llewellyn Smith, Nucl. Phys. {\bf B317} 526 (1989).
\bibitem{Smith}
N. Isgur and C.H. Llewelly Smith, Phys. Rev. Let. {\bf }52 1080 (1984).
A.V. Radyushkin, Acta Phys. Polonica, B {\bf 15} 403 (1984).
A.P. Bakulev and A.V. Radyushkin, Phys. Lett. B {\bf 271} 223 (1991).
\bibitem{Shen}
T. Huang and Q.X. Shen, Z. Phys. C {\bf 50} 139 (1991).
\bibitem{Botts}
J. Botts and G. Sterman, Nucl. Phys. {\bf B325} 62 (1989).
\bibitem{Li}
H.N. Li and G. Sterman, Nucl. Phys. {\bf B381} 129 (1992).
\bibitem{Li2}
H.N. Li, Phys. Rev. D {\bf 48} 4243 (1993).
\bibitem{Jakob}
R. Jakob and P. Kroll, Phys. Lett. B {\bf 315} 463 (1993).
\bibitem{}
J. Bloz, R. Jakob, P. Kroll and M. Bergmann 
Phys. Lett. B {\bf 342} 345 (1995).
R. Jakob, Phys. Rev. D {\bf 50} 5647 (1994).
\bibitem{Akhoury}
R. Akhoury, G. Sterman, and Y.P. Yao, Phys. Rev. D {\bf 50} 358 (1994).
\bibitem{cao}
F.G. Cao, T. Huang and C. W. Luo, Phys. rev D {\bf 52} 5358 (1995).
\bibitem{Land}
P.V. Landshoff, Phys. Rev. D {\bf 10} 1024 (1974).
\bibitem{CFG}
F.G. Cao, B.Q. Ma, and T. Huang, in preparation.
\bibitem{BHL}
S. J. Brodsky, T. Huang and G. P. Lepage, in {\it particles and
fields-2}, Proceedings of the Banff Summer Institute, Banff, Alberta, 
1981, edited by A. Z. Capri and A. N. Kamal (Plenum, New York,1983), P. 143;
G. P. Lepage, S. J. Brodsky, T. Huang, and P. B. Mackenize, {\it ibid.}, P. 83;
T. Huang, in {\it Proceedings of XXth International Conference on High Energy 
Physics}, Madison, Wisconsin, 1980, edited by L. Durand and L. G. Pondrom, 
Aip Con. Proc. No. 69 (AIP, New York, 1981), p. 1000.
\bibitem{MABQ}
T. Huang, B. Q. Ma, and Q. X. Shen, Phys. Rew. D {\bf 49} 1490 (1994).
\bibitem{Huang1}
T. Huang, in {\it Proceedings of the International Symposium 
on Particle and Nuclear physics}, Beijing, 1985, edited by N. Hu and
C. S. Wu (World Scientific, Singapore, 1986), p. 151; in {\it Proceedings of 
the Second Asia Pacific Physics Conference}, India, 1986 (Word Scientific, 
Singapore, 1987), p. 258.
\bibitem{Huang2}
T. Huang, in {\it High $p_{t}$ Physics and Higher Twists},
Proceeding of the Workshop, Pairs, France, 1988, edited by M. Benayuon,
M. Fontannaz, and J. L. Narjoux [Nucl. Phys. B (Proc. Suppl.) 
{\bf 7B} (1989) 320.
\bibitem{Dzi}
Z. Dziembowski and L. Mankiewicz, Phys. Rev. Lett. {\bf 58} 2175 (1987);
Z. Dziembowski, Phys. Rev. D {\bf 37} 778 (1987).
\end{thebibliography}
\end{document}